# Site-Selective Oxygen Vacancy Formation Derived from the Characteristic Crystal Structures of Sn-Nb Complex Oxides

*Akane Samizo[1,2,\*], Makoto Minohara[2], Naoto Kikuchi[2], Kyoko K. Bando[3], Yoshihiro Aiura[2], Ko Mibu[4], and Keishi Nishio[1]*

[1]Department of Materials Science and Technology, Tokyo University of Science, Katsushika 125-8585, Japan

[2]Research Institute for Advanced Electronics and Photonics, National Institute of Advanced Industrial Science and Technology (AIST), Tsukuba, Ibaraki 305-8568, Japan

[3]Nanomaterials Research Institute, National Institute of Advanced Industrial Science and Technology (AIST), Tsukuba, Ibaraki 305-8565, Japan

[4]Department of Physical Science and Engineering, Nagoya Institute of Technology, Nagoya 466-8555, Japan

**ABSTRACT:** Divalent tin oxides have attracted considerable attention as novel *p*-type oxide semiconductors, which are essential for realizing future oxide electronic devices. Recently, *p*-type $Sn_2Nb_2O_7$ and $SnNb_2O_6$ were developed; however, enhanced hole mobility by reducing defect concentrations is required for practical use. In this work, we investigate the correlation between the formation of oxygen vacancy ($V_O^{\cdot\cdot}$), which may reduce the hole-generation efficiency and hole mobility, and the crystal structure in Sn-Nb complex oxides. Extended X-ray absorption fine structure spectroscopy and Rietveld analysis of x-ray diffraction revealed the preferential formation of $V_O^{\cdot\cdot}$ at the O site bonded to the Sn ions in both the tin niobates. Moreover, a large amount of $V_O^{\cdot\cdot}$ around the Sn ions were found in the *p*-type $Sn_2Nb_2O_7$, thereby indicating the effect of $V_O^{\cdot\cdot}$ to the low hole-generation efficiency. The dependence of the formation of $V_O^{\cdot\cdot}$ on the crystal structure can be elucidated from the Sn-O bond strength that is evaluated based on the bond valence sum and Debye temperature. The differences in the bond strengths of the two Sn-Nb complex oxides are correlated through the steric hindrance of $Sn^{2+}$ with asymmetric electron density distribution. This suggests the importance of the material design with a focus on the local structure around the Sn ions to prevent the formation of $V_O^{\cdot\cdot}$ in *p*-type $Sn^{2+}$ oxides.

**INTRODUCTION**

Oxide semiconductors with high electrical conductivity and transparency have attracted significant attention owing to their application in several technologies such as thin-film solar cells and/or touch screens. However, most of the existing oxide semiconductors are *n*-type oxides, such as Sn-doped $In_2O_3$ (ITO) and $InGaO_3(ZnO)_5$ (IGZO) [1], and the lack of practical *p*-type oxide semiconductors limits their applications to unipolar devices. Fabrications of the *p-n* junctions, as same as technologies for Si semiconductors, will lead to more energy saving, and more complex transparent devices, and thus it is important to develop *p*-type oxides for the future "oxide electronics" [2].

One of the intrinsic problems in the development of *p*-type oxide semiconductors is the low hole mobility due to a flat valence band maximum composed of an O 2p orbital [3, 4]. To reduce strong localization of holes, modifying the valence band structure by hybridization of the O 2p orbital with a metallic s or d orbital is theoretically suggested [3]. Because the s orbital has an electron density with a large spatial spread and strong interactions with the neighboring atoms, the degree of valence band dispersion is particularly large for the s orbital-based materials. Thus, oxides including cations with $ns^2$ electronic configuration (such as $Sn^{2+}$, $Pb^{2+}$, and $Bi^{3+}$) have attracted attention as new candidates for *p*-type oxides [3-7]. This approach was demonstrated through the fabrication of SnO films with a hole mobility of ~21 $cm^2V^{-1}s^{-1}$ [8-11]. However, because SnO has an indirect band gap of 0.7 eV [9], it is not suitable for future transparent device applications. For realizing both high hole mobility and transparency, *p*-type tin oxides with wide band gaps are required.

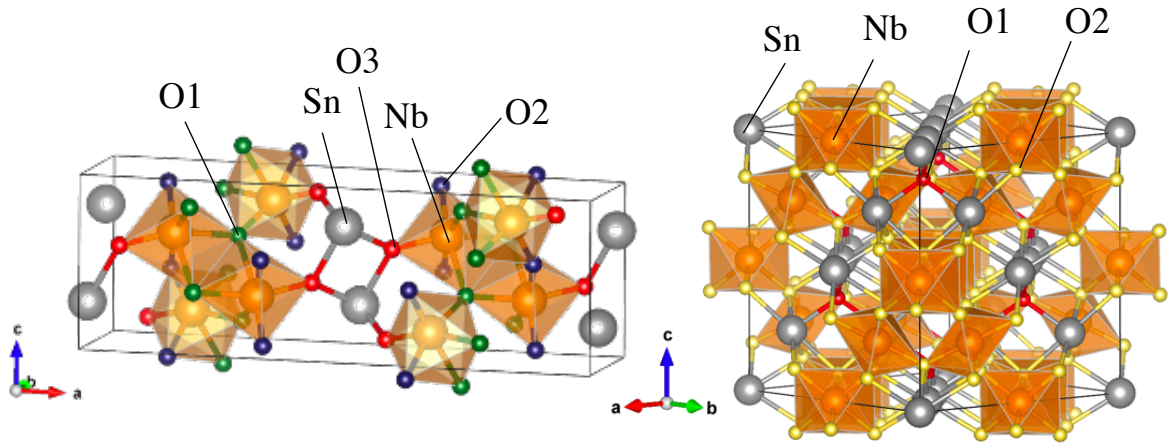

**Figure 1.** Crystal structures of $SnNb_2O_6$ (left) and $Sn_2Nb_2O_7$ (right). The information for each site is provided in previous reports [13,14]. The O ions are denoted in different color based on the site. The closest O site to Sn (O3 and O1 for $SnNb_2O_6$ and $Sn_2Nb_2O_7$, respectively) is denoted in red.

Recently, we developed novel *p*-type $Sn^{2+}$ oxides, $SnNb_2O_6$ and $Sn_2Nb_2O_7$, with a band gap of 2.4 eV (Fig. 1) [13,14]. Although their valence band is composed of Sn 5s orbital, they showed low hole mobility of 0.03–0.4 $cm^2V^{-1}s^{-1}$ so far [12-14]. In general, since structural defects act as the scattering centers for electrons (holes), the mobility declines with the increase of the defect density, and hence, the reduction of these defects is a common strategy for improving hole mobility [18]. By the same time, the carriers are generated by the defect formations, therefore, a high carrier generation efficiency of the donor (acceptor) is required to improve the hole mobility for practical applications. Thus, it is important to focus on the underlying mechanism of the defect formations. The above-mentioned Sn-Nb complex oxides are known to have three native defects: the Sn vacancy ($V''_{Sn}$), the $Sn^{4+}$ substitutional defect at the $Nb^{5+}$ site ($Sn'_{Nb}$), and oxygen vacancy ($V^{\cdot\cdot}_O$) [19-24]. The charge carriers are generated by the defect formations, as shown below,

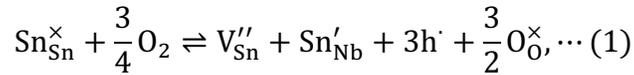

$$Sn^{\times}_{Sn} + \frac{3}{4}O_2 \rightleftharpoons V''_{Sn} + Sn'_{Nb} + 3h^{\cdot} + \frac{3}{2}O^{\times}_O, \cdots (1)$$

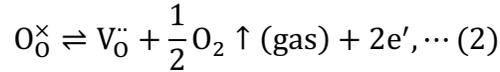

$$O^{\times}_O \rightleftharpoons V^{\cdot\cdot}_O + \frac{1}{2}O_2 \uparrow (gas) + 2e', \cdots (2)$$

where $Sn^{\times}_{Sn}$ and $O^{\times}_O$ are $Sn^{2+}$ at the Sn site and $O^{2-}$ at the O site. These defect formations occur simultaneously during sample fabrication. Since the Eqs. (1) and (2) generate holes and electrons, respectively, the balance of these defect formations determines the carrier type and density. In fact, the carrier type of these Sn-Nb complex oxides can be manipulated by controlling the annealing conditions [12,13]. The net carrier density ($n_h$ and $n_e$ for hole and electron densities, respectively) and $Sn'_{Nb}$ density ($N_{Sn'Nb}$) of $SnNb_2O_6$ and $Sn_2Nb_2O_7$ for both the *p*- and *n*-types are summarized in Table 1. With the change in the carrier type from *p*-type to *n*-type, the amount of $Sn'_{Nb}$ clearly decreases. Here, we note that the amount of net carrier of *p*-type samples cannot be quantitatively described by the amount of $Sn'_{Nb}$ ; the hole densities are two or three orders magnitude lower than $Sn'_{Nb}$ density. The degree of such mismatch between hole and $Sn'_{Nb}$ density of $SnNb_2O_6$ is different from that of $Sn_2Nb_2O_7$. When the hole-generation efficiency ($\eta_{Hole}$) is defined by $n_h / 3N_{Sn'Nb}$ based on Eq. (1), the $\eta_{Hole}$ value of $SnNb_2O_6$ is found to be three orders of magnitude higher than that of $Sn_2Nb_2O_7$. Considering the fact that the Sn-Nb complex oxides are annealed in reductive atmosphere to prevent oxidative decompositions, it is naturally expected that the charge compensation was occurred by the formation of large amounts of $V_O^{\cdot\cdot}$ as shown in Eqs. (2). Thus, in order to clarify the mechanism of appearing *p*-type carriers in electrical properties, the $V_O^{\cdot\cdot}$ formation has to be considered. In this study, we investigated the formation of $V_O^{\cdot\cdot}$ in $SnNb_2O_6$ and $Sn_2Nb_2O_7$ through extended X-ray absorption fine structure (EXAFS) spectroscopy. We found that $V_O^{\cdot\cdot}$ can be formed preferentially around the Sn ions for both $SnNb_2O_6$ and $Sn_2Nb_2O_7$. In addition, it was revealed that the degree of the structural disorders which derived from the $V_O^{\cdot\cdot}$ formation was higher in p-type $Sn_2Nb_2O_7$ with low $\eta_{Hole}$ than that in the *p*-type $SnNb_2O_6$. These findings can be well explained based on the Sn-O bond strength in the two different crystal structures.

**Table 1** Hole densiry ($n_h$), electron density($n_e$), $Sn'_{Nb}$ density ($N_{Sn'Nb}$), and hole-generation efficiencies ($\eta_{Hole}$) in $SnNb_2O_6$ and $Sn_2Nb_2O_7$. The ϖalue of $N_{Sn'Nb}$ of n-type $SnNb_2O_6$ is lower than the $Sn^{4+}$ detection limit of Mössbauer spectroscopy.

|  | Carrier type | $n_h$ ($n_e$) [cm$^{-3}$] | $N_{Sn'Nb}$ [cm$^{-3}$] | $\eta_{Hole}$ |
|---|---|---|---|---|
| $SnNb_2O_6$ | *p*-type | $3.7 \times 10^{18}$ | $8.6 \times 10^{19}$ | $1.4 \times 10^{-2}$ |
|  | *n*-type | $7.5 \times 10^{15}$ | 0.0 | - |
| $Sn_2Nb_2O_7$ | *p*-type | $2.5 \times 10^{17}$ | $1.7 \times 10^{21}$ | $4.9 \times 10^{-5}$ |
|  | *n*-type | $4.9 \times 10^{15}$ | $7.3 \times 10^{20}$ | - |

## EXPERIMENTAL METHOD

**Sample Preparation**

$Sn_2Nb_2O_7$ and $SnNb_2O_6$ were prepared through solid-state reactions. The starting materials used were SnO (Kojundo Chemical Laboratory; purity, 99.5 %) and $Nb_2O_5$ (Kojundo Chemical Laboratory; purity, 99.9 %). They were mixed in an agate mortar with ethanol and dried in air for 24 h. The mixed powder was then calcined at 1173 K in an alumina tube furnace under $N_2$ at a flow rate of 150 ml min$^{-1}$. The synthesized samples were annealed to obtain *p*-type and *n*-type conductivities as follows. The calcined powder was ground again in an agate mortar, followed by mixing with a polyvinyl alcohol (PVA) aqueous solution (PVA: 2 wt.% to the sample) and ethanol. After air-drying for 24 h, the dried samples were sieved to obtain narrow sieve fractions < 212 μm. The obtained powder was pressed isostatically at 290 MPa to form discs of 12 mm diameter. The sample disks were annealed in the tube furnace under each of following heating conditions: 1023 K under a $N_2$ atmosphere at a flow rate of 50 ml min$^{-1}$ for *p*-type $Sn_2Nb_2O_7$ and $SnNb_2O_6$, 1473 K under the $N_2$ atmosphere at a flow rate of 150 ml min$^{-1}$ for *n*-type $Sn_2Nb_2O_7$, and 1473 K under the $N_2$ atmosphere at a flow rate of 100 ml min$^{-1}$ for *n*-type $SnNb_2O_6$.

**Structural Characterization**

The crystal structure was confirmed by X-ray diffraction (XRD), which was conducted using the Bragg–Brentano configuration with Cu Kα radiation (PANalytical, X'Pert Pro MPD). The structure parameters of $Sn_2Nb_2O_7$ were refined by the Rietveld analysis using the RIETAN-FP software [25]. The crystal structure was obtained by the VESTA software [26].

The local crystal structures were investigated by EXAFS measurements. The EXAFS spectra of all the samples were measured by the transmission mode at 40 K. The Sn K-edge and Nb K-edge spectra were observed at the beamline AR-NW10A of the Photon Factory Advanced Ring (PF-AR) and BL-9C of Photon Factory, KEK [27]. The EXAFS spectra were Fourier-transformed using the Hanning window function within the $k$ range of 3.0–14.0 Å$^{-1}$ for the Sn K-edge, Nb K-edge of SnNb$_2$O$_6$, and Nb K-edge of Sn$_2$Nb$_2$O$_7$, and 3.0–13.0 Å$^{-1}$ for the Sn K-edge of Sn$_2$Nb$_2$O$_7$. The data processing of the EXAFS spectra was performed using Athena [28].

The $^{119}$Sn Mössbauer spectroscopy was conducted by the conventional transmission method under the same experimental geometry at 78 and 300 K. The powder samples were mixed with silicone grease and plastered on pure Al foil to eliminate any orientation effect. The range of the Doppler velocity of the Ca$^{119m}$SnO$_3$ source was set to ±8 mm s$^{-1}$. The peak position was calibrated using a CaSnO$_3$ reference.

**RESULTS AND DISCUSSIONS**

**Comparing the formation of oxygen vacancies through crystal structure analysis**

Before discussing the possible $V_O^{\cdot\cdot}$ formation and its crystal structure dependency, we provide an experimental evidence of the comparable crystalline quality between the SnNb$_2$O$_6$ and Sn$_2$Nb$_2$O$_7$ samples. Figures S1(a) and (b) show the XRD patterns of the *p*- and *n*-type samples (see supplementary section S1). The Bragg peaks assigned for SnNb$_2$O$_6$ (JCPDS : 98-020-2827) and Sn$_2$Nb$_2$O$_7$ (JCPDS : 98-027-9575) were observed to be sharp. This indicates sufficient crystalline quality to perform the EXAFS spectroscopy and discuss the possible local disorder in the crystal structure due to the formation of $V_O^{\cdot\cdot}$.

The Fourier transforms (FT) of the Sn K-edge and Nb K-edge EXAFS spectra of SnNb$_2$O$_6$ are shown in Figs. 2 (a) and (b), respectively. In Fig. 2(a), the peak between 1.5 and 1.7 Å in the FT of the Sn K-edge EXAFS spectra corresponds to the first-neighbor Sn-O bond. The peak intensity of the Sn-O bond in the *n*-type sample is distinctly weaker than that in the *p*-type sample. Here, the spectral intensities of the FT feature mainly reflect the coordination number ($N_j$) of the absorbing atom and the Debye Waller factor ($\sigma_j$) of the scattering atom, which corresponds to the magnitude of thermal vibrations depending on the crystalline quality. Considering the higher annealing temperature of the *n*-type sample than the *p*-type one ($|\Delta T| = 450$ K), $\sigma_j$ of the *n*-type sample is naturally expected to be lower than that of the *p*-type one. Namely, the FT magnitude of the *n*-type sample could be stronger than that of the *p*-type one; however, this contradicts the experimental findings. Therefore, the significant decrease in the Sn-O peak intensity can be assumed to be a decrease in $N_j$, that is, formation of $V_O^{\cdot\cdot}$. Simultaneously, we notice that the EXAFS oscillation becomes unclear for the *n*-type sample (inset of Fig. 2(a)), suggesting the formation of structural disorder around the Sn ions.

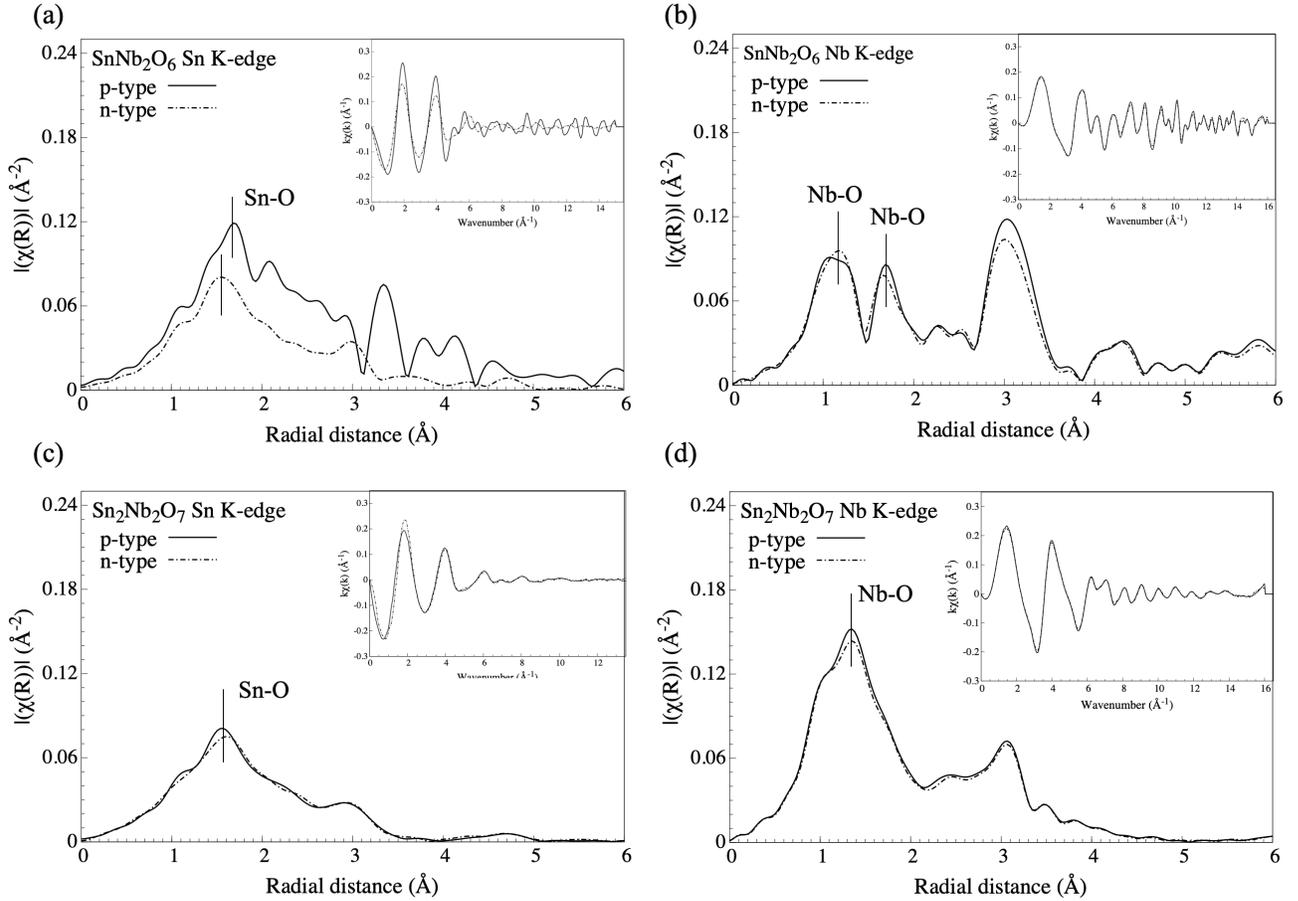

**Figure 2**. Extracted EXAFS oscillations (inset) and the associated k-weighted Fourier transform of (a) Sn K-edge, (b) Nb K-edge of $SnNb_2O_6$, and (c) Sn K-edge, (d) Nb K-edge of $Sn_2Nb_2O_7$. The *p*- and *n*-type samples are denoted by solid and dotted lines, respectively.

In contrast, in Fig. 2 (b), two peaks are observed in the FT of the Nb K-edge EXAFS spectra for radial distance (*r*) < 2 Å. However, a slightly complex behavior is observed in the FT spectra of the Nb K-edge compared to that of the Sn K-edge; the intensity of the first shell (*r* = 1.1 Å) for the *n*-type sample seems to be stronger than that for the *p*-type one; however, the second shell (*r* = 1.7 Å) shows an opposite behavior. As mentioned above, the FT magnitude of the *n*-type sample will be higher than that of *p*-type one, when we assume a low $\sigma_j$ values with higher annealing temperatures for the *n*-type samples. Thus, this complex feature could reflect the decrease in $N_j$. It is noteworthy that the decrease in the intensity is significant for the Sn K-edge as compared to Nb K-edge. From the spectral calculations using *FEFF* program [29, 30], the first metal-oxide scattering peaks in the Sn K-edge ($r$ ~1.5 Å) and Nb K-edge ($r$ ~1.8 Å) EXAFS spectra mainly originate from the Sn-O3 bonds and Nb-O1 and Nb-O2 bonds, respectively (see supplementary S2). Given the large decrease in the intensity of the FT EXAFS spectra for the Sn K-edge, we can consider that $V_O^{\cdot\cdot}$ is preferentially formed at O3 sites rather than at O1 and/or O2 sites. The site-selective $V_O^{\cdot\cdot}$ formation is consistent with the findings in the Rietveld analysis of the XRD patterns (see supplementary S3).

Figures 2 (c) and (d) show the FT of the Sn K-edge and Nb K-edge EXAFS spectra of $Sn_2Nb_2O_7$, respectively. The intensity of the *n*-type sample for both the FT spectra is slightly weaker than that of the *p*-type one, suggesting larger formation of $V_O^{\cdot\cdot}$ for the *n*-type sample. Initially, negligible differences are observed between the Sn K-edge and Nb K-edge; the preferential formation of $V_O^{\cdot\cdot}$ for *n*-type is not evident for $Sn_2Nb_2O_7$ as it is for $SnNb_2O_6$. However, upon closer look at the EXAFS oscillations at wavenumber ($k$) > 6 Å, Sn K-edge EXAFS oscillation is unclear for both *p*-type and *n*-type $Sn_2Nb_2O_7$(inset of Fig. 2(c)), which is similar to the behavior observed for *n*-type $SnNb_2O_6$. This suggests that the local structural disorder occurs around the Sn ions. In the crystal structure of $Sn_2Nb_2O_7$, there are two different oxygen sites named O1 and O2, where the former and latter correspond to the oxygen sites closest to Sn and Nb, respectively. Considering its structural analogy to $SnNb_2O_6$, the findings from EXAFS oscillations suggest that there are large amounts of $V_O^{\cdot\cdot}$ at O1 in $Sn_2Nb_2O_7$.

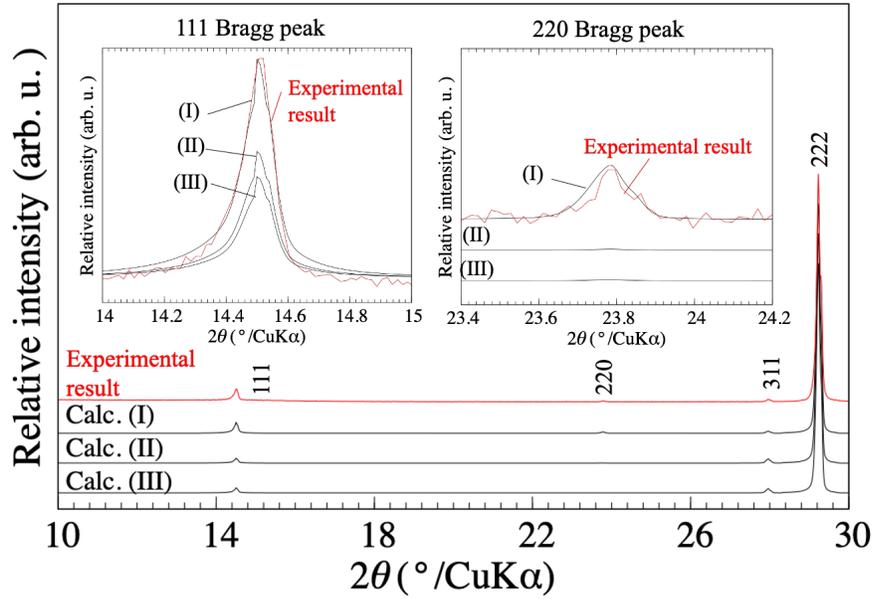

**Figure 3.** Comparison of experimental result and calculated XRD patterns of $p$-type $Sn_2Nb_2O_7$ based on the (I) O1-deficient model, (II) ideal model without $V_O^{\cdot\cdot}$, and (III) O2-deficient model. The insets show the magnification around 111 (left, $2\theta = 14.0$–$15.0°$) and 220 (right, $2\theta = 23.4$–$24.2°$) Bragg peaks. The intensity was normalized as the relative intensity, which is the ratio of the intensities of each Bragg peak to the highest 222 peak intensity.

Again, the Rietveld analysis supports the large $V_O^{\cdot\cdot}$ formation at O1 site in $Sn_2Nb_2O_7$ [14]. Figure 3 shows a comparison of the experimental result with the calculated XRD patterns of $p$-type $Sn_2Nb_2O_7$ (the structural parameters for the calculated patterns are shown under section S4). The intensity is normalized by the intensity of the 222 Bragg peak. Here, the calculated patterns (I), (II), and, (III) indicate the O1-deficient model, ideal model without $V_O^{\cdot\cdot}$ formation, and O2-deficient model, respectively. The insets show the magnification around the 111 ($2\theta = 14.0$–$15.0°$) and 220 ($2\theta = 23.4$–$24.2°$) Bragg peaks. As for the 111 Bragg peak, the O1-deficient model shows the highest intensity. Moreover, the 220 Bragg peak can be clearly observed only in pattern (I). These characteristics in pattern (I) are consistent with the experimental result, which indicates the site-selective $V_O^{..}$ formations at the O1 site. Therefore, the structural disorder, which is potentially triggered by the $V_O^{..}$ formation, occurs preferentially around the Sn ions. These findings are supported by our preliminary calculations on the defect formation energy [16].

Further, we focus on the difference between $p$-type $SnNb_2O_6$ and $p$-type $Sn_2Nb_2O_7$. As shown in Figs. 2(a) and (c), the peak intensity of Sn-O bonds in FT spectra for $p$-type $Sn_2Nb_2O_7$ is lower than that for $p$-type $SnNb_2O_6$ and comparable with that of oxygen-deficient $n$-type $SnNb_2O_6$. This can also be observed in the EXAFS oscillations as mentioned above. These findings strongly suggest large amount of $V_O^{..}$ formation around the Sn ions in not only for $n$-type $Sn_2Nb_2O_7$ but also $p$-type one. Therefore, poor $\eta_{Hole}$ in $p$-type $Sn_2Nb_2O_7$ occurs due to the larger amount of $V_O^{..}$ formation than in $p$-type $SnNb_2O_6$.

**Strength of Sn-O bond in Sn-Nb complex oxides**

To understand the different $V_O^{..}$ formations around the Sn ions between $SnNb_2O_6$ and $Sn_2Nb_2O_7$, we note the Sn-O bond strength in each structure. Bond valence sum (BVS) is a suitable indicator of the bond strength [31]. The bond valence $S_{ij}$ is directly related to the strength of the Sn-O bond and inversely with the bond length [32,33]. It can be approximated as

$$S_{ij} = \exp\left(\frac{R_0 - R_{ij}}{B}\right), \quad \cdots (3)$$

where $R_0$ is the bond valence parameter, 1.984 Å for Sn-O bond and 1.916 Å for Nb-O bond, $R_{ij}$ and $B$ are the actual bond length and 0.37 which is a constant independent of an element, respectively [34, 35]. The BVS is the sum of $S_{ij}$, which is expressed as follows [36].

$$\text{BVS} = \sum_j S_{ij}. \qquad \cdots (4)$$

The values of the BVS for the Sn and O of $SnNb_2O_6$ and $Sn_2Nb_2O_7$ are summarized in Table 2. The $R_{ij}$ (< 3 Å) values, which were evaluated from the crystal structure data of the *p*-type samples refined by the Rietveld analysis, are summarized in Table S3 (see supplementary section S5). The BVS of the Sn of $SnNb_2O_6$ is higher than that of $Sn_2Nb_2O_7$, suggesting that the Sn and O are tightly bonded in $SnNb_2O_6$ as compared to that in $Sn_2Nb_2O_7$.

Table 2. Bond valence sum (BVS) of the $SnNb_2O_6$ and $Sn_2Nb_2O_7$.

| SnNb$_2$O$_6$ | | Sn$_2$Nb$_2$O$_7$ | |
|---|---|---|---|
| atom | BVS | atom | BVS |
| O1 | 1.90 | O1 | 1.57 |
| O2 | 1.95 | O2 | 1.93 |
| O3 | 1.89 | | |
| Sn | 2.37 | Sn | 1.54 |

Another indicator for the Sn-O bond strength is Debye temperature ($\theta_D$), which is based on the amplitude of thermal vibration [37]. We estimated $\theta_D$ of $Sn^{2+}$ for *p*-type $SnNb_2O_6$ and $Sn_2Nb_2O_7$ by means of $^{119}$Sn Mössbauer spectroscopy. Based on the high-temperature approximation of the Debye model, the Debye temperature is expressed with recoil-free fractions $f(T)$ as follows [38],

$$\ln f(T) = {-6E_R T}/{k_B \theta_D^2}, \qquad \cdots (5)$$

$$A = \text{const.} \times f, \qquad \cdots (6)$$

where $A$, $\theta_D$, $T$, $k_B$, and $E_R$ are the integral absorption intensity of Mössbauer peak, Debye temperature, measurement temperature, Boltzmann constant, and recoil energy, respectively. The temperature dependence of ln $A$ and the estimated values of $\theta_D$ are shown in Fig. 4 (details are provided in supplementary section S6). The slope of $Sn_2Nb_2O_7$ was larger than that of $SnNb_2O_6$; Therefore, the $\theta_D$ value of $SnNb_2O_6$ (237 K) was higher than that of $Sn_2Nb_2O_7$ (174 K). This is qualitatively in good agreement with the results obtained through the BVS estimations.

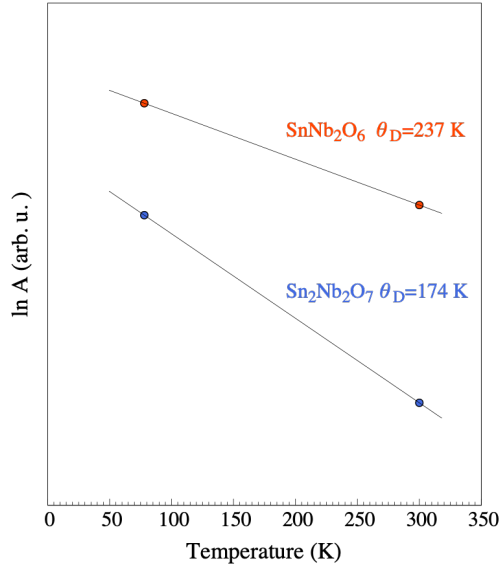

**Figure 4.** Temperature dependence of ln $A$ obtained through the $^{119}$Sn Mössbauer spectra and Debye temperature $\theta_D$ of $Sn^{2+}$ sites for $SnNb_2O_6$ and $Sn_2Nb_2O_7$.

**Comparing local structures around the Sn ions**

Finally, we will discuss the origin of the differences in the Sn-O bond strengths of $SnNb_2O_6$ and $Sn_2Nb_2O_7$ by focusing on the local structures around the Sn ions. It is known that Sn $5s^2$ electrons have an asymmetric electron density and a large volume as same as that of oxygen ions [23]. Moreover, they act like the lone electron pair in the $Sn^{2+}$ oxides [39,40]. The polarization of the $5s^2$ electrons occurs due to the hybridization of Sn 5s with O 2p and Sn 5p orbitals [39]. The schematic images of the local structures around the Sn ions of $SnNb_2O_6$ and $Sn_2Nb_2O_7$ are shown in Fig. 5. In the case of SnNb$_2$O$_6$, eight oxygen ions are arranged at the vertices of a square antiprism with the Sn ion at the center. The distance between Sn and O2 is large (>3 Å); therefore, there is a large space present toward the O2 side. The Sn 5s$^2$ electrons can exist in space with a small steric hindrance. However, in the "ideal pyrochlore" structure of Sn$_2$Nb$_2$O$_7$, Sn is surrounded by eight closely arranged oxygen ions. To form an ideal pyrochlore structure, the defects and/or disorders are necessary [19-24] because large electrostatic repulsions occur between the Sn 5s$^2$ electrons and oxygen ions. Considering the characteristic crystal structure geometry of Sn$_2$Nb$_2$O$_7$, the $V_O^{\cdot\cdot}$ formation at the O1 site could provide the space for the asymmetric electron density of Sn$^{2+}$. This led to a large amount of $V_O^{\cdot\cdot}$ formation for Sn$_2$Nb$_2$O$_7$, particularly around the Sn ions. Therefore, we can interpret the reason for the lower hole-generation efficiency of Sn$_2$Nb$_2$O$_7$ as compared to SnNb$_2$O$_6$ based on the characteristic crystal geometry.

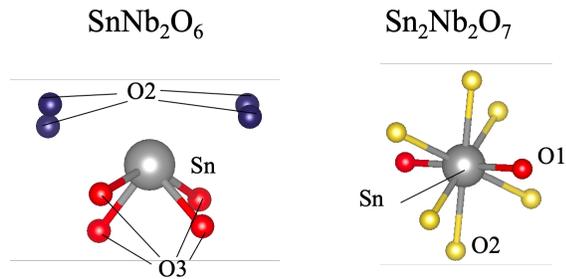

**Figure 5** Schematic image of the local structure around the Sn ions of SnNb$_2$O$_6$ (left) and Sn$_2$Nb$_2$O$_7$ (right).

**CONCLUSIONS**

In conclusion, we investigated the oxygen vacancy ($V_O^{\cdot\cdot}$) formation in *p*-type $SnNb_2O_6$ and $Sn_2Nb_2O_7$ to reveal the origin of the differences in the hole-generation efficiencies $\eta_{Hole}$. The EXAFS spectra and Rietveld analysis indicated the preferential $V_O^{\cdot\cdot}$ formation at the oxygen sites, which were bonded to Sn ions in both the Sn-Nb complex oxides. Moreover, a larger amount of $V_O^{\cdot\cdot}$ formation was observed in $Sn_2Nb_2O_7$ with low $\eta_{Hole}$ as compared to $SnNb_2O_6$ with high $\eta_{Hole}$. The difference in the $V_O^{\cdot\cdot}$ formations was due to the Sn-O bond strength, which is correlated with the local structure of $Sn^{2+}$ and the asymmetric electron density of the Sn $5s^2$ electrons. Conversely, the distorted structure around the Sn ions and the long Sn-O bond length, such as those of $SnNb_2O_6$, resulted in higher $\eta_{Hole}$ through the suppression of $V_O^{\cdot\cdot}$ formation. Because such a structure has a large space for the Sn $5s^2$ electrons, it reduced the large electrostatic repulsions between the Sn $5s^2$ electrons and oxygen ions. To date, the development of *p*-type oxide which has *s*-orbital based VBM (*s*-orbital-based *p*-type oxide) has been limited because of the difficulty in controlling the formations of $V_O^{\cdot\cdot}$. Our data suggest the importance of the local structure around the Sn ions to prevent the $V_O^{\cdot\cdot}$ formations in the *p*-type $Sn^{2+}$-based complex oxides. Moreover, this approach is potentially adapted for other oxides with n$s^2$ cations (such as $Pb^{2+}$ and $Bi^{3+}$) because they have similar n$s^2$ electrons with the asymmetric electron density and large volume. The findings in this study give us new direction for exploring a candidate of *s*-orbital-based *p*-type oxides through a reduction of the scattering center generated by structural defect formations.

**ASSOCIATED CONTENT**

**Supporting Information**

The supporting information is available free of charge on the ACS Publication website.

XRD patterns of $SnNb_2O_6$ and $Sn_2Nb_2O_7$, *FEFF* calculations of Sn K-edge and Nb K-edge EXAFS of $SnNb_2O_6$, Summary of Rietveld analysis of *n*-type $SnNb_2O_6$, Structural parameters of $Sn_2Nb_2O_7$ of O1 and O2 deficient model for calculated XRD patterns, Sn-O bond length of *p*-type $SnNb_2O_6$ and $Sn_2Nb_2O_7$, Mössbauer spectra and calculation of Debye temperature (PDF)

**AUTHOR INFORMATION**

**Corresponding Author**

Akane Samizo − Tokyo University of Science, Katsushika, Japan; orcid.org/0000-0002-7488-385X; E-mail : a-samizo@rs.tus.ac.jp

**Author**

Makoto Minohara − National Institute of Advanced Industrial Science and Technology (AIST), Tsukuba, Japan; orcid.org/0000-0003-4367-9175

Naoto Kikuchi − National Institute of Advanced Industrial Science and Technology (AIST), Tsukuba, Japan; orcid.org/0000-0003-1035-3515

Kyoko K. Bando − National Institute of Advanced Industrial Science and Technology (AIST), Tsukuba, Japan

Yoshihiro Aiura − National Institute of Advanced Industrial Science and Technology (AIST), Tsukuba, Japan; orcid.org/0000-0002-4478-7680

Ko Mibu− Nagoya Institute of Technology, Nagoya, Japan; orcid.org/0000-0002-6416-1028

Keishi Nishio− Tokyo University of Science, Katsushika, Japan

**Author contributions**

A.S., N.K., K.N. planned the whole experiments. A.S., M.M., N.K., K.K.B., Y.A. carried out

EXAFS measurements. A.S., N.K., K.N. performed Rietveld analysis and XRD measurements. K.M. and A.S. carried out [119]Sn Mössbauer spectroscopy measurements and performed the analysis of the spectra. A.S. wrote the paper with input from all co-authors.

**Notes**
The authors declare no competing financial interest.

## ACKNOWLEDGMENTS

This work was supported by a Grant-in-Aid for Scientific Research (20K22472, 18K05285) from the Japan Society for the Promotion of Science (JSPS), and Tokyo University of Science (TUS) Grant for Young and Female Researchers. The work at KEK-PF was performed under the approval of the Program Advisory Committee (Proposals No. 2018G685, 2019G070, 2019G543, 2020G657) at the Institute of Materials Structure Science, KEK. We also thank Dr. Hiroaki Nitani for his technical support on the EXAFS measurements. The Mössbauer measurements were conducted at Nagoya Institute of Technology, supported by the Nanotechnology Platform Program (Molecule and Material Synthesis) of the Ministry of Education, Culture, Sports, Science and Technology (MEXT), Japan. We would like to thank Editage (www.editage.com) for English language editing.

## REFERENCES

[1] Nomura, K.; Ohta, H.; Takagi, A.; Kamiya, T.; Hirano, M.; Hosono, H. Room-temperature fabrication of transparent flexible thin-film transistors using amorphous oxide semiconductors. Nature, 2004, 432, 488–492.

[2] Coll, M.; et al. Towards Oxide Electronics: A Roadmap. Appl.Surf. Sci. 2019, 482, 1–93.

[3] Kawazoe, H.; Yanagi, H.; Ueda, K.; Hosono, H. Transparent *p*-Type Conducting Oxides: Design and Fabrication of *p-n* Heterojunctions. MRS Bull. 2000, 25, 28-36.

[4] Hautier, G.; Miglio, A.; Ceder, G.; Rignanese, G.-M.; Gonze, X. Identification and Design Principles of Low Hole Effective Mass *P*-type Transparent Conducting Oxides. Nat. Commun. 2013, 4, 2292.

[5] Zhang, K. H. L.; Xi, K.; Blamire, M. G.; Egdell, R. G. *P*-type Transparent Conducting Oxides. J. Phys.: Condens. Matter 2016, 28, 383002.

[6] Wang, Z.; Nayak, P. K.; Caraveo-Frescas, J. A.; Alshareef, H. N. Recent Developments in p-Type Oxide Semiconductor Materials and Devices. Adv. Mater. 2016, 28, 3831–3892.

[7] Bhatia, A.; Hautier, G.; Nilgianskul, T.; Miglio, A.; Sun, J.; Kim,H. J.; Kim, K. H.; Chen, S.; Rignanese, G. M.; Gonze, X.; Suntivich, J.High-Mobility Bismuth-based Transparent *p*-Type Oxide from High-Throughput Material Screening. Chem. Mater. 2016, 28, 30–34.

[8] Ogo, Y.; Hiramatsu, H.; Nomura, K.; Yanagi, H.; Kamiya, T.; Hirano, M.; Hosono, H. *P*-Channel Thin-Film Transistor Using *P*-type Oxide Semiconductor, SnO. Appl. Phys. Lett. 2008, 93, 032113.

[9] Ogo, Y.; Hiramatsu, H.; Nomura, K.; Yanagi, H.; Kamiya, T.; Hirano, M.; Hosono, H. Tin Monoxide as an s-Orbital-Based P-type Oxide Semiconductor: Electronic Structures and TFT Application. Phys. Status Solidi A 2009, 206, 2187-2191.

[10] Minohara, M.; Kikuchi, N.; Yoshida, Y.; Kumigashira, H.; Aiura,Y. Improvement of the hole mobility of SnO epitaxial films grown by pulsed laser deposition. J. Mater. Chem. C 2019, 7, 6332–6336.

[11] Minohara, M.; Samizo A.; Kikuchi, N.; Bando K. K.; Yoshida Y.; Aiura,Y. Tailoring the Hole Mobility in SnO Films by Modulating the Growth Thermodynamics and Kinetics. J. Phys. Chem. C. 2020, 124, 1755-1760.

[12] Kikuchi, N.; Samizo, A.; Ikeda, S.; Aiura, Y.; Mibu, K.; Nishio, K. Carrier Generation in a P-type Oxide Semiconductor: $Sn_2(Nb_{2-x}Ta_x)O_7$. Phys. Rev. Mater. 2017, 1, 021601.

[13] Samizo, A.; Kikuchi, N.; Aiura, Y.; Nishio, K.; Mibu, K. Carrier Generation in p-Type Wide-Gap Oxide: $SnNb_2O_6$ Foordite. Chem. Mater. 2018, 30, 8221–8225.

[14] Samizo, A.; Kikuchi, N.; Nishio, K. Effect of Crystal Structure on Hole Carrier Generation in Wide-gap P-type Tin-Niobate. MRS Adv. 2019, 4, 1, 27-32.

[15] Hosogi, Y.; Shimodaira, Y.; Kato, H.; Kobayashi, H.; Kudo, A. Role of $Sn^{2+}$ in the Band Structure of $SnM_2O_6$ and $Sn_2M_2O_7$ (M = Nb and Ta) and Their Photocatalytic Properties. Chem. Mater. 2008, 20, 1299–1307.

[16] Aiura Y.; Ozawa K.; Hase I.; Bando K. K.; Haga H.; Kawanaka H.; Samizo A.; Kikuchi N.; Mase K. Disappearance of Localized Valence Band Maximum of Ternary Tin Oxide with Pyrochlore Structure, $Sn_2Nb_2O_7$. J. Phys. Chem. C 2017,121, 9480-9488.

[17] Katayama, S.; Hayashi, H.; Kumagai, Y.; Oba, F.; Tanaka, I. Electronic Structure and Defect Chemistry of Tin(II) Complex Oxide $SnNb_2O_6$. J. Phys. Chem. C 2016, 120, 9604–9611.

[18] Chattopadhyay, D.; Queisser, H. J. ; Electron Scattering by Ionized Impurities in Semiconductors. Rev. Mod. Phys., 1981, 53, 745-768.

[19] Stewart, D. J; Knop, O.; , Meads, R. E.; Parker, W. G.; Pyrochlores. IX. Partially Oxidized $Sn_2Nb_2O_7$, and $Sn_2Ta_2O_7$: A Mössbauer Study of Sn(I1,IV) Compounds. Can. J. Chem. 1973, 51, 1041-1049.

[20] Subramanian, M. A.; Aravamudan, G.; Subba, Rao, G. V.; OXIDE PYROCHLORES – A REVIEW. Prog. Solid State Chem. 1983, 15, 55-143.

[21] Birchall, T.; Sleight, A. W.; Nonstoichiometric Phases in the Sn-Nb-O and Sn-Ta-O Systems Having Pyrochlore-Related Structures. J. Solid State Chem. 1975, 13, 118-130.

[22] Mizoguchi, H.; Wattiaus, A.; Kykyneshi, R.; Tate, J.; Sleight, A. W.; Subramanian, M. A. Synthesis and Characterization of $Sn^{2+}$ Oxides with the Pyrochlore Structure. Mater. Res. Bull. 2008, 43, 1943-1948.

[23] Cruz, L. P.; Savariault, J. -M.; Rocha, J. Pyrochlore-Type Tin Niobate. Acta Cryst. Sect. C, 2001, 57, 1001-1003.

[24] Cruz, L. P.; Savariault, J. -M.; Rocha, J.; Jumas, J. -C.; Pedrosa de Jesus J. D., Synthesis and Characterization of Tin Niobates. J. Sol. St. Chem. 2001, 156, 349-354.

[25] Izumi, F.; Momma, K. Three-Dimensional Visualization in Powder Diffraction. Sol. St. Phenom. 2007, 130, 15–20.

[26] Momma, K.; Izumi, F. VESTA 3 for Three-Dimensional Visualization of Crystal, Volumetric and Morphology Data J. Appl. Crystallogr. 2011, 44, 1272-1276.

[27] Nomura, M.; Koike, Y.; Sato, M.; Koyama, A.; Inada, Y.; Asakura, K. A New XAFS Beamline NW10A at the Photon Factory. AIP Conf. Proc. 2007, 882, 896.

[28] Ravel, B.; Newville, M. ATHENA, ARTEMIS, HEPHAESTUS: Data Analysis for X-ray Absorption Spectroscopy Using IFEFFIT. J. Synchrotron Radiat. 2005, 12, 537–541.

[29] Zabinsky, S. I.; Rehr, J. J.; Ankudinov, A.; Albers R. C.; Eller M. J. Multple-Scattering Calculations of X-ray-Absorption Spectra, Phys. Rev. B 1995, 52, 2995.

[30] Ankudinov, A. L.; Ravel, B.; Rehr, J. J.; Conradson, S. D. Real-Space Multiple-Scattering Calculation and Interpretation of X-ray-Absorption Near-Edge Structure, Phys. Rev. B 1998, 58, 7565.

[31] Brown, I. D.; Shannon, R.D.; Empirical Bond-Strength-Bond-Length Curves for Oxides. Acta Cryst. 1973, A29, 266-282.

[32] Wang, X.; Liebau, F.; Influence of Lone-Pair Electrons of Cations on Bond-Valence Parameters. Z. Kristallogr. 1996, 211, 437-439.

[33] Wang, X. The Contribution To Bond Valences By Lone Electron Pairs. Mater. Res. Soc. Symp. Proc. 2005, 848, FF7.4.1.

[34] Brese, N. E.; O'Keeffe, M., Bond-Valence Parameters for Solids. Acta Cryst B. 1991, 47, 192-197.

[35] Tytko, K. H.; Mehmke, J.; Kurad D. Bond Length-bond Valence Relationships with Particular Reference to Polyoxometalate Chemistry. In: Bard A.J. et al. (eds) Bonding and Charge Distribution in Polyoxometalates: A Bond Valence Approach. Structure and Bonding, vol 93. 1999, Springer, Berlin, Heidelberg.

[36] Camargo, P.H.C.; Brown, I. D. The Chemical Bond in Inorganic Chemistry: the Bond Valence Model, 2nd ed. J. Mater. Sci. 2017, 52, 9959–9962.

[37] Jiang, D.; Wu, M.; Liu, D.; Li, F.; Chai, M.; Liu, S. Structural Stability, Electronic Structures, Mechanical Properties and Debye Temperature of Transition Metal Impurities in Tungsten: A First-Principles Study. Metals 2019, 9, 967.

[38] Gütlich, P. H.; Link, R.; Trautwein, A. Mössbauer Spectroscopy and Transition Metal Chemistry. 1978, Springer-Verlag, Berlin.

[39] Walsh, A.; Payne, J. D.; Egdell G. R.; Watson, W. G.; Stereochemistry of Post-Transition Metal Oxides: Revision of the Classical Lone Pair Model. Chem. Soc. Rev. 2011, 40, 4455-4463.

[40] Galy, J.; Meunier, G.; Anderson, S.; Åatröm, A.; Stéréochimie des eléments comportant des paires non liées: Ge (II), As (III), Se (IV), Br (V), Sn (II), Sb (III), Te (IV), I (V), Xe (VI), Tl (I), Pb (II), et Bi (III) (oxydes, fluorures et oxyfluorures). J. Sol. St. Chem. 1975, 13, 142-159.

Insert Table of Contents artwork here

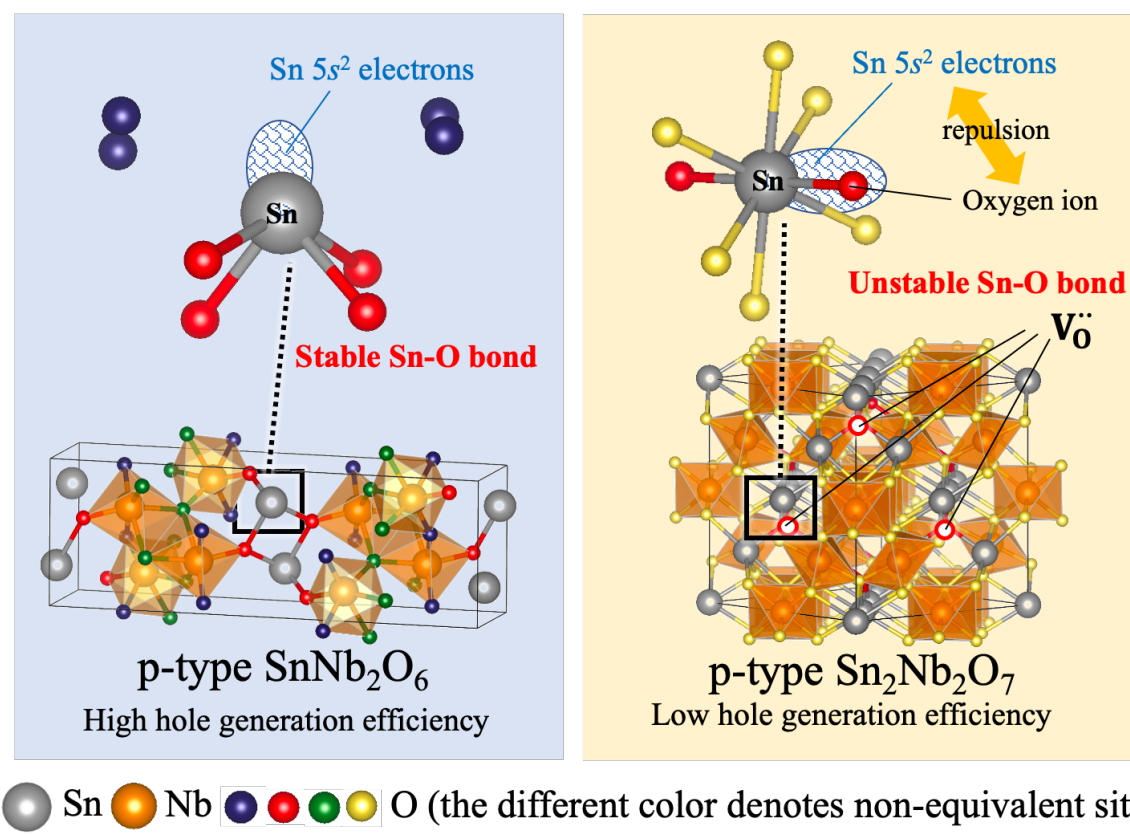